\documentclass[referee]{aa} 
\usepackage{graphicx}
\usepackage{txfonts}
\usepackage{natbib}
\bibpunct{(}{)}{;}{a}{}{,}

\begin{document}
\title{Atomic Radiative Data for Oxygen and Nitrogen for Solar Photospheric Studies}

\author{Manuel A.~Bautista\inst{\ref{inst1}}\and Maria Bergemann\inst{\ref{inst2}}\and Helena Carvajal Gallego\inst{\ref{inst3}}\and S\'ebastien Gamrath\inst{\ref{inst3}}\and Patrick Palmeri\inst{\ref{inst3}}\and Pascal Quinet\inst{\ref{inst3},\ref{inst4}}}

\institute{Department of Physics, Western Michigan University, Kalamazoo,
MI 49008, USA \email{manuel.bautista@wmich.edu}\label{inst1}
\and 
Max Planck Institute for Astronomy, 69117, Heidelberg, Germany\label{inst2}
\and
Physique Atomique et Astrophysique, Universit\'e de UMONS, B-7000 Mons, Belgium\label{inst3}
\and 
IPNAS, Universit\'e de Li\`ege, B-4000 Li\`ege, Belgium\label{inst4}
}

\date{Received <date> /
Accepted <date>}

\abstract {} 
{Our recent re-analysis of the solar photospheric spectra with non-local thermodynamic equilibrium (non-LTE) models resulted in higher metal abundances compared to previous works. When applying the new 
chemical abundances to Standard Solar Model calculations, the new composition resolves the long-standing discrepancies with independent constraints on the solar structure from helioseismology.}
{Critical to the determination of chemical abundances is the accuracy of the atomic data, specially the $f$-values, used in the radiative transfer models. Here we describe in detail the calculations of $f$-values for neutral oxygen and nitrogen used in our non-LTE models.}
{Our calculations of $f$-values are based on a multi-method, multi-code approach and are the most detailed and extensive of its kind for the spectral lines of interest. We also report in this paper the details of extensive R-matrix calculation of photo-ionization cross sections for oxygen.} 
{Our calculation resulted in reliable $f$-values with well constrained uncertainties.
We compare our results with previous theoretical and experimental determinations {of atomic data. We also quantify the influence of adopted photo-ionisation cross-sections on the spectroscopic estimate of the solar O abundance, using the data from different sources. We confirm that the 3D non-LTE value presented by Bergemann et al. 2021 is robust and unaffected by the choice of photo-ionisation data, contrary to the recent claim made by Nahar.} 
}{}


\titlerunning{$f$-values for {stellar spectroscopy}}
\authorrunning{Butista et al.}

\maketitle

\section{Introduction}

Among elements heavier than helium, the abundances of carbon (C), nitrogen (N) and oxygen (O) amount to about 2/3 of all elements by mass. Moreover, O is the major contributor to the opacity in the solar interior that determines the transport of energy and thus the interior structure of the Sun.
Determining accurate abundances of these elements is critically important to understand the solar structure and evolution. 
In addition, measurements of CNO abundances in stars are crucial to address a variety of different problems in modern astrophysics, including but not limited to studies of Galactic chemical evolution and enrichment by asymptotic giant branch stars \citep{bensby,bergemann18,schuler}, 
multiple populations in stellar clusters \citep{2022A&A...658A..80T}, structure of proto-planetary discs \citep{2021ApJ...909...40T}, planet formation and planetary atmospheres 
\citep{2021NatAs.tmp..253B, 2022NatAs...6..226B}. 

The most reliable solar chemical abundances are obtained from the solar photospheric spectrum \citep{lodders}.  Regarding the solar O abundance, various groups have studied it in different ways.
\citet{grevesse} deduced an O photospheric abundance of 8.83$\pm$0.06 dex, based on 1D LTE methods. By contrast, recent estimates from detailed non-LTE analysis with 3D radiation-hydrodynamics (RHD)
simulations of convection, are log A(O) = 8.69 $\pm$ 0.04 dex (Asplund et et al. 2021), and 
log A(O) = 8.76$\pm$0.07 dex (Caffau et al. 2008). The former abundance leads to solar interior opacities that are too low to satisfy the standard solar model calculations \citep{serenelli2009, serenelli2011,villante}. 
This is because these opacities lead to a predicted internal structure of the present-day Sun,
with sound speed profile, depth of the convective envelope, and surface He abundancen,
which is contradicted by independent constraints on its structure obtained from helioseismology.

\citet{bergemann} re-analyzed the solar photospheric O abundance using revised and more 
complete atomic data, non-LTE calculations with 1D hydrostatic (MARCS) and 3D 
radiation-hydrodynamics and magneto-hydrodynamics simulations of the solar atmosphere fron the {\sc STAGGER}  
\citep{stagger} and {\sc BIFROST} \citep{bifrost} codes. the Bergemann et al.  analysis also takes into account
the influence of the chromosphere. By comparing the theoretical line profiles with high-resolution, R$\approx$  700~000,
spatially-resolved spectra of the Sun, they found log A(O) = 8.74 $\pm$ 0.03 dex from the 
permitted O~I lines at 777 nm and log A(O) = 8.77$ \pm$ 0.05 dex from the forbidden [O~I] line at 630 nm.
In \citet{magg}, we extended this approach to estimate the abundances of all important 
chemical elements in the solar photosphere. This includes a new estimate of the N abundance based on newly calculated oscillator strengths. The revised abundances are higher compared to those reported in previous non-LTE analysis by \citet{2021A&A...653A.141A}. 
Our revised values yield solar interior opacities in much better agreement with the helioseismic constraints.

Key parameters in all chemical abundance determinations are the radiative transition probabilities adopted, as errors in the adopted A-values directly affect the derived abundances. The vast majority of A-values come from theoretical calculations and the differences in results published by different authors often exceed the reported uncertainties. Uncertainties in calculated values can be broadly categorized as statistical and systematic. The former shall include the dispersion in numerical results arising from different treatments of the atomic potential and the optimization techniques of atomic orbitals. The systematic uncertainties are much more difficult to predict and to quantify, because they result from incomplete descriptions of atomic states. Such descriptions of states are generally done by configuration interaction (CI) and/or level mixing schemes that leads to expansions, which must be truncated by computational limitations.

In this paper, we present the details of the calculations of radiative transition probabilities for N and O, which contributed to a significant revision of the solar chemical abundances of these elements. We are especially concerned with the uncertainties of the derived transitions rates. 
To this end, we have carried out large, exhaustive calculations of $f$-values for the particular O and N lines measured in the observed solar spectra. For these calculations we use three different and independent theoretical methods and compare the results in detail.

As we were preparing this publication, \citet{nahar} criticized the atomic data used in \citet{bergemann} and put in doubt our new 3D non-LTE estimate of the solar oxygen abundance. Nahar's criticism focuses specially on the photoionization cross sections of O I states. Thus, we include here a detailed description of our cross sections. Further, we demonstrate how the differences in the cross-sections influence the solar spectroscopic models and resulting solar O abundance using independent sources of photoionization cross sections.

\section{Transition probabilities and $f$-values}

In this section we describe the calculation of radiative transition probabilities. We used four different 
computer codes employing different theoretical approaches. These are: (i) {\sc AUTOSTRUCTURE} \citep{autoss} that is based on the Thomas-Fermi-Dirac-Amaldi central potential;
(ii) the R-matrix method impletented in the RMATRX package \citep{rmatrx}; 
(iii) the HFR+CPOL approach based on 
the pseudo-relativistic Hartree-Fock (HFR) method initially introduced by \cite{cowan} and modified to take 
core-polarization effects into account \citep{quinet99,quinet02}; and (iv) the 
{\sc GRASP2018} code \citep{grasp18} based on the fully relativistic multiconfiguration Dirac-Hartree-Fock (MCDHF) method 
\citep{grant,grasp18}. 

In providing the most accurate radiative rates and $f$-values for transitions of O I and N I, we also need to provide reliable uncertainty estimates for these rates. The specific lines of interest are, for oxygen: $\lambda\lambda$7771.94, 7774.42, 7775.54 \AA (2p$^3$3s~$^5$S$^o_2$ ~--~2p$^3$3p~$^5$P$_{3,2,1}$); and for nitrogen:
$\lambda$8629.235 \AA (2p$^2$3s~$^2$P$_{3/2}$ ~-~2p$^2$3p~$^2$P$^o_{3/2}$),
$\lambda$8683.403 \AA (2p$^2$3s~$^4$P$_{3/2}$ ~-~2p$^2$3p~$^4$D$^o_{5/2}$).

\subsection{{\sc AUTOSTRUCTURE} Calculations}

{\sc AUTOSTRUCTURE} solves the Breit--Pauli
Schr\"odinger equation with a scaled Thomas--Fermi--Dirac (TFD) central-field potential to generate orthonormal 
atomic orbitals. Configuration Interaction (CI) atomic state functions are built using these orbitals. 
The scaling factors for each orbital are optimized in a multiconfiguration variational procedure
minimizing a sum over LS term energies or a weighted average of LS term energies. Perturbative
corrections (Term Energy Correction; TEC) are applied to the multi-electronic Hamiltonian, which adjusts theoretical LS 
term energies closer the centers of gravity of the available experimental multiplets. A final empirical correction (Level Energy Correction; LEC) is applied to reproduce level energy separations when computing $gf$-values and transitions probabilities.

We carry out numerous calculations, systematically adding additional configurations to the CI expansion. In doing so, we observe the evolution of the $gf$-values as they reach convergence and we can quantify statistical and systematic uncertainties in the final rates. We aim at including, within computational limits, all configurations that contribute to the rates for the transitions of interest, while maintaining good overall representation for the entire atomic system.

It should be noted that many configurations in the CI expansion that contribute to a particular transition rate may not 
influence the energies of the levels involved in any significant way. Thus, one cannot generally assess the importance of including specific configurations to the transitions of interest based only on the predicted level energies. Instead, one needs to carry out the full calculation, including orbital optimizations, TECs and LECs, up to computing the actual radiative rates.
This makes the work presented here computationally intensive. Orbital optimization often requires numerous calculations of the atomic model. After each calculation one must compare in detail the predicted energies with experimental values, compute differences and find the term dentification numbers and level densitifcation numbers assigned by the code, which change from model to model. Once the potential is satisfactorily optimized one uses the difference between predicted and experimental term averaged energies for a new calculation with TECs.  At this 
stage, the LECs are produced from the difference with experiment in fine-structure level energies, and a final large calculation of all transition rates is carried out. This procedure has to be automatized, because of the very large datasets involved, and in order to minimise errors caused by inspection by eye. Therefore, a Python code was used, which semi-automatically generates {\sc AUTOSTRUCTURE} input files, reads the output files, compares
with experimental data, and produces TEC and LEC factors.
%
%Doing all of the tasks described above, and even reading the correct transitions rates from output files of several GBs in size is a very lengthy and prompt to error process if done manually. Instead, we wrote a Python code running on Anaconda Notebooks, which semi-automatically generates {\sc AUTOSTRUCTURE} input files, reads the output files, makes comparisons with experimental data, and produces TEC and LEC factors.

We start the computations by selecting the orbitals that need to be accounted for in our calculations. By inspection of results obtained with CI expansion of the form 2s$^2$2p$^3nl$ for 
oxygen and 2s$^2$2p$^2nl$ for nitrogen, we find that only orbitals with $n\le 5$ and $l\le 3$ 
for oxygen and $n\le 6$ and $l\le 4$ for nitrogen make significant contributions to the transitions of interest. These orbitals define finite sets of possible configurations to be included in our CI expansions. However, all together these amount to thousands of configurations for each atom and including them all at once would exceed presently available computational resources. Therefore, our approach is to start from a relatively small CI expansion that yields a reasonably good representation of levels up to those of interest. Then, we progressively add sets of configurations by single-, double- and triple-electron promotions. At each stage we retain only those new configurations that affect the transition rates of interest and discard unimportant configurations. This way, we seek to obtain the most accurate rates possible while keeping the CI expansion manageable. Below we present details of the calculations for each species and the results.

\subsubsection{Oxygen}
We start with a 15-configuration expansion we refer to as CI-A. This expansion includes orbitals up to principal quantum
number $n=4$ and angular momentum $l\le 2$, as shown in Table~\ref{oiconfigs}. This expansion gives a very reasonable structure in terms of energies (see Table~\ref{oiener}). The theoretical energies shown in this table are {\it ab initio} 
energies, prior to term and level energy corrections. One can see that the CI-A expansion gives acceptable energies 
(within $\sim 10\%$ from experiment) for terms of the 2p$^4$ configuration. The energies of terms of the 2p$^3$3s and 
2p$^3$4s configurations are very good. The expansion also gives good energies for the lowest terms of the 2p$^3$3p and 2p$^3$3d configurations, but overestimate significantly the energies of the higher terms. This is because the angular momentum splitting of these configurations is inaccurate in the {\it ab initio} representation.
Nonetheless, the CI-A expansion gives term energies within $\sim 2\%$ of experiment for the states 2p$^3$3s~$^5$S$^o$ 
and 2p$^3$3p~$^5$P. Moreover, the representation of all terms is expected to be improved by using TECs.

\begin{figure}
\scalebox{0.5}{\includegraphics[angle=-90]{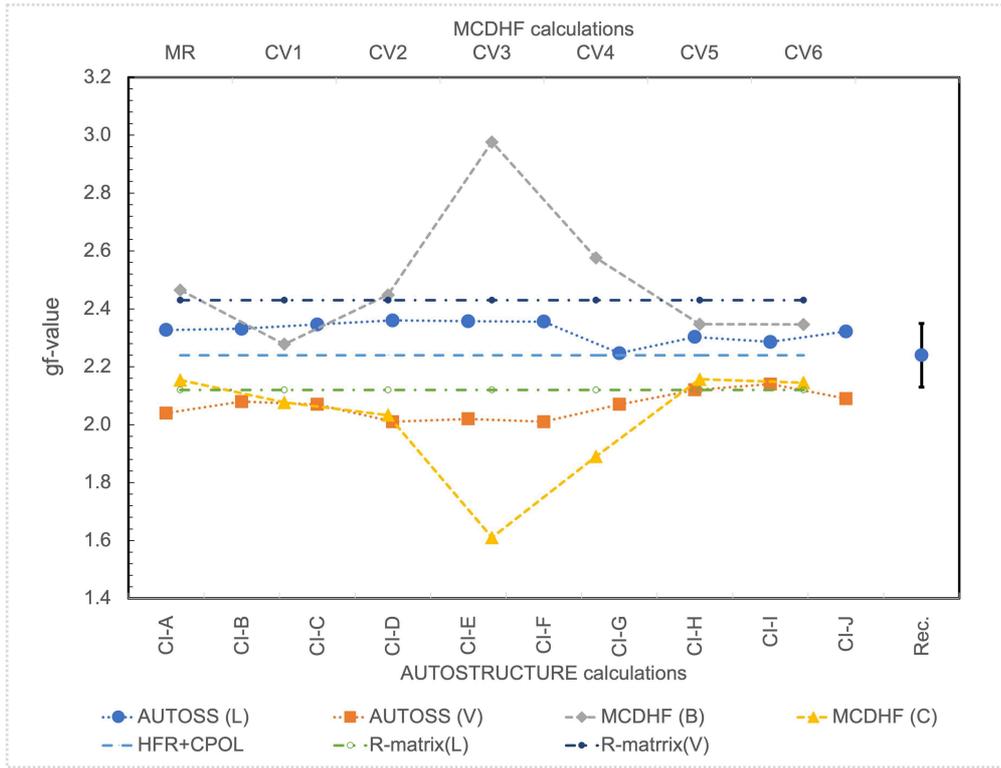}}
  \caption{$gf$-values for the O~I $\lambda$7771 line. The
results of all different model calculations are shown increasing order of complexity from left to right.
The values depicted are {\sc AUTOSTRUCTURE} length (AUTOSS (L)) and velocity (AUTOSS (V)) gauges as circles and 
square points, repectively, MCDHF Babushkin (MCDHF (B)) and Coulomb (MCDHF (C)) gauges by diamond and triangular points, respectively, HFR+CPOL by horizonal long-dashed line, and R-matrix length (R-matrix(L)) and velocity (R-matrix(V)) gauges by dot-dashed lines.
The point with error bars at the right end of the plot depicts our recommended $gf$-value for the transitions and its
estimated uncertainty. \label{OAdipole}} 
\end{figure}

Expansion CI-B extends the original expansion by one electron promotions from the 2s orbital. Expansion CI-C extends CI-B by configurations with two electrons in $n=4$.
These expansions change neither the atom's description nor the transition rates of interest (see Fig.~\ref{OAdipole}) in a significant way.

Expansions CI-D and CI-E extend the original expansion by two electron promotions from 2s. Again, this results in only small changes to the atomic structure and changes to the transition rates of less than 1\%. If anything, these expansions lead to slight increases in the difference between length and velocity gauges of the $gf$-values of the dipole transitions.

In expansions CI-F, CI-G and CI-H we explore additional two- and three-electron promotions from 2s and 2p orbital to $n=3$ and 4 valence orbitals. In all cases, the energies remain essentially unchanged and so are the transitions rates.

Expansions CI-I and CI-J, with 49 and 106 configurations respectively, are the largest we ran. Both of these include $n=5$ orbitals. In CI-I we include configurations with up to three electron promotions, one electron from 2s and up to two electrons from 2p.
In CI-J we have configurations with up to four electron promotions.
The additional configurations yield no large changes to the predicted atomic
structure, but the difference between length and velocity gauges of the
$gf$-values is reduced from $\sim 15\%$ to $\sim 9\%$.

Figure~\ref{OAdipole} shows the $gf$-values for the $\lambda$7771 dipole allowed line. 
Other lines of this multiplet have the same behavior, thus we do not show them.
Both length and velocity gauges of the  $gf$-values as obtained with {\sc AUTOSTRUCTURE} are shown in the figure. TECs and LECs have been included in all calculations. Without these the scatter in the obtained rates is spuriously larger than shown in the figure. 
The results our R-matrix, HFR+CPOL and MDCHF calculations are also shown here and they are discussed below.
It can be seen the $gf$-values for this transition are remarkably stable and the length and 
velocity gauges agree well (within $\sim 10\%$). We conclude the transition rate is converged and reliable. 
%
%
%The behavior of the forbidden transition is illustrated in Figure~\ref{OAforbidden}. As expected, the scatter in this rate is a lot greater than for the dipole transitions. Various expansions yield significantly different A-values while giving very similar level energies.
%Nonetheless, it is safe to conclude the rate is close to 0.012 s$^{-1}$ and the uncertainty, based on the dispersion among various results, is estimated at $\sim 15\%$.

\subsubsection{Nitrogen}

The nitrogen transitions of interest here converge much more slowly than the oxygen lines. 
To achieve reliable $gf$-values with {\sc AUTOSTRUCTURE}  we had to progressively increase the size of the CI expansion 
to include up to 355 configurations. The different expansions used here are described in Table~\ref{niconfigs} and the resultant, uncorrected energies from these expansions are presented in Table~\ref{niener}.

All expansions yield acceptable term energies, within $\sim$10\% of the experimental values, with the exception of energies for the 2s$^2$2p$^3$~$^2$P$^o$ term from the CI-A and CI-B expansions. These two expansions overestimate the energy split among terms of the ground configuration 2s$^2$2p$^3$. Unaccounted electron-electron correlations are at fault here, but this is resolved by adding configurations with two-electron promotions from the 2s orbital. This is done in CI-C and all larger expansions. For all expansions we find the predicted energies for excited configurations are systematically underestimated. This is due to strong relaxation effects of the 2p orbital. This is likely to introduce error in the calculated transition rates, though TECs will mitigate these errors in sufficiently large CI expansions.

When computing radiative rates and $gf$-values we find significant discrepancies between the length and velocity gauges. The velocity gauge values, which preferentially weights the small radius part of the wavefunctions, is often less reliable than the length gauge. Thus, we consider the former set of values for the uncertainty estimates.

\begin{figure}
\scalebox{0.5}{\includegraphics[angle=+90]{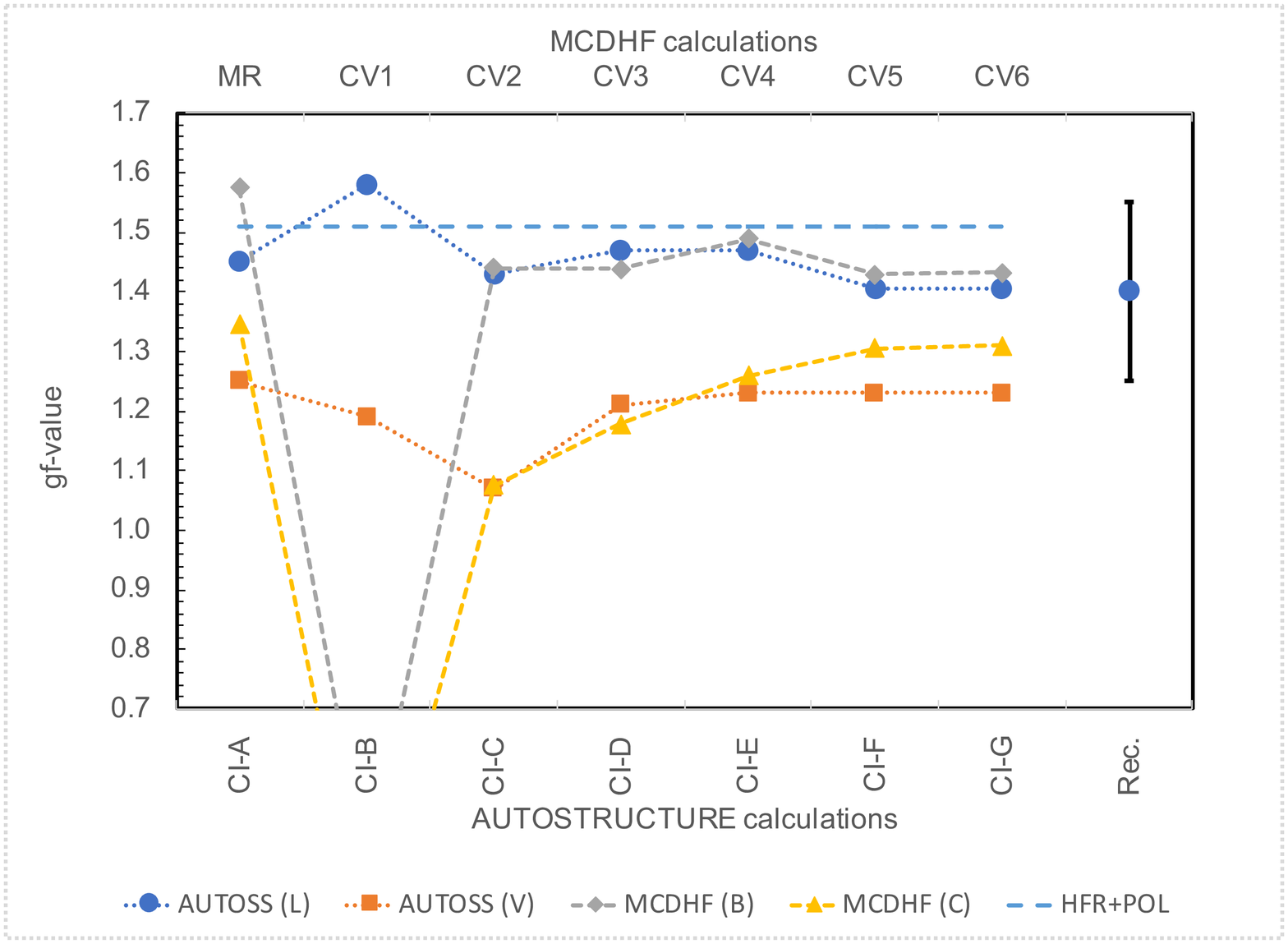}}
  \caption{$gf$-values for the N~I $\lambda 8683$ line. 
The different point types and lines are as in Figure 1. \label{Ndipole1}}
\end{figure}

\begin{figure}
\scalebox{0.5}{\includegraphics[angle=+90]{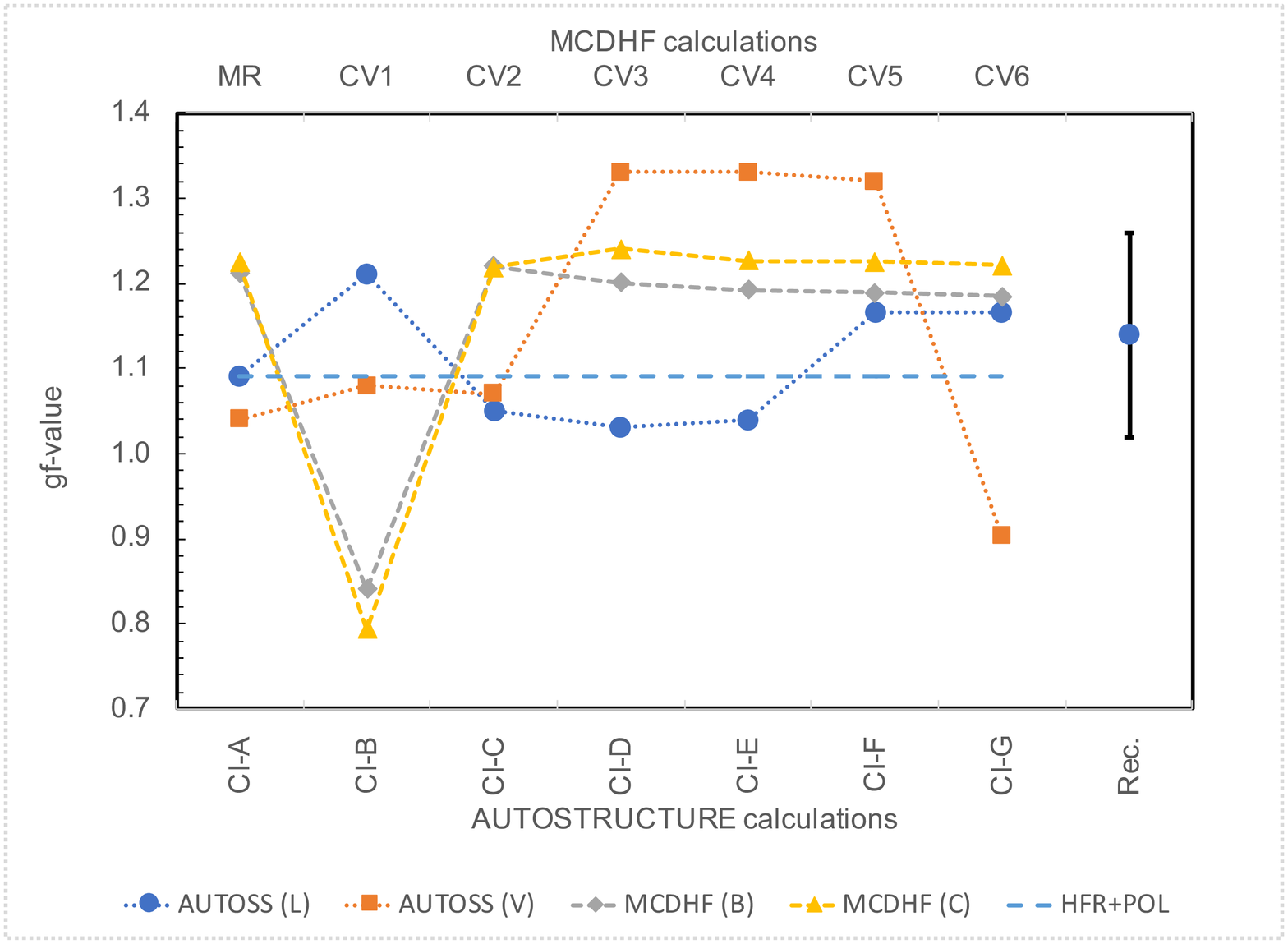}}
  \caption{$gf$-values for the N~I $\lambda 8629$ line. 
The different point types and lines are as in Figure 1. \label{Ndipole2}}
\end{figure}

Figures~\ref{Ndipole1}, \ref{Ndipole2} depict the evolution of the $gf$-values for the N I $\lambda 8683$ and $\lambda 8629$ lines, respectively. We see the value for the $\lambda 8683$ line is scattered between 1.4 and 1.6, while the scatter for the $\lambda 8629$ lines ranges from 1.0 to 1.2. From the minimum expansion CI-A to CI-B we see both gf-values to jump up by more than 10\%, as a result of two-electron promotions from the 2p orbital. Then, two-electron promotions from the 2s, included in expansion CI-C bring the gf-values down again. Simultaneous promotions from 2s and 2p, included in configurations CI-D and CI-E seem to bring the values to converge to about 1.47 and 1.03, respectively. 
However, adding $n$=5 orbitals in expansions CI-F and CI-G changes the gf-values again to 1.40 and 1.16. These last expansions, with 126 and 355 configurations, yield nearly identical results and we expect them to have reached convergence.

\subsection{R-matrix}

The R-matrix method allows for the calculation of bound level energies and $gf$-values for 
electric dipole transitions.
Here, the wavefunctions of the bound levels are constructed as CI expansions of basis state
functions built using the close coupling approximation. This method is implemented in the
RMTRX computer package \citep{rmatrx}.

We computed oxygen $gf$-values for the transitions of inerest from the same calculation carried out for 
photoionization cross sections.
The detail of this calculation are presented in section 4.
As verification of the quality of the calculations, Table~\ref{oihfrenergy}
compares the {\it ab-initio} predicted bound energies of the levels of interest with experimental energy values
compiled by NIST \citep{nist}. It can be seen the predicted energy of the 3s~$^5$S$_2$ levels is underestimated by about 6\%, while the energies of 3p~$^5$P$_j$ 
levels are within 5\% of experiment. This is considered good accuracy for predicted bound state energies and provides support of the reliability of the results.

R-matrix calculations of $gf$-values have the advangante of including a lot of CI very 
efficiently. The down side, though, is that the clculations are too envolved to allow for 
detailed follow up of the convergence of specific transitions.

\subsection{Pseudo-relativistic Hartree-Fock + Core-Polarization Calculations}

Another approach used to compute oscillator strengths in O I and N I was the HFR+CPOL 
method. This consists of the pseudo-relativistic Hartree-Fock (HFR) method \citep{cowan} and 
later modified to take core-polarization effects into account 
\citep{quinet99, quinet02}. The computational strategies developed for each atomic system are detailed next.

ables~\ref{oihfrenergy} and \ref{nihfrenergy}
compare the calculated energies of the levels of interest in oxygen and nitrogen, respectively, with experimental energies
from NIST \citep{nist}. The details of the calculations are described next.

\subsubsection{Oxygen}

In the case of neutral oxygen, the valence-valence interactions were considered among a set of configurations including
2p$^4$, 2p$^3nl$ (with $nl$ up to 6h) and all single and double (SD) excitations from 2s, 2p to 3s, 3p, 3d orbitals 
while the core-valence interactions involving the 1s subshell were modeled by a core-polarization potential. 
More precisely, the 2s$^2$2p$^4$, 2s$^2$2p$^3$$n$p ($n$ = 3--6), 2s$^2$2p$^3$$n$f ($n$ = 4--6), 2s$^2$2p$^3$6h, 2s$^2$2p$^2$3s$^2$, 2s$^2$2p$^2$3p$^2$, 2s$^2$2p$^2$3d$^2$, 2s$^2$2p$^2$3s3d, 2s$^2$2p3p$^3$, 2s$^2$2p3s$^2$3p,2s$^2$2p3p3d$^2$, 2s$^2$2p3s3p3d, 2s2p$^4$3s, 2s2p$^4$3d, 2s2p$^3$3s3p, 2s2p$^3$3p3d, 2p$^6$, 2p$^5$3p, 2p$^4$3s$^2$, 2p$^4$3p$^2$, 2p$^4$3d$^2$, 2p$^4$3s3d, 2p$^3$3p$^3$, 2p$^3$3s$^2$3p, 2p$^3$3p3d$^2$, 2p$^3$3s3p3d even-parity configurations, and the 2s$^2$2p$^3$$n$s ($n$ = 3--6), 2s$^2$2p$^3$$n$d ($n$ = 3--6), 2s$^2$2p$^3$$n$g ($n$ = 5--6), 2s$^2$2p$^2$3s3p, 2s$^2$2p$^2$3p3d, 2s$^2$2p3d$^3$, 2s$^2$2p3s$^2$3d, 2s$^2$2p3s3p$^2$, 2s$^2$2p3p$^2$3d, 2s$^2$2p3s3d$^2$, 2s2p$^5$, 2s2p$^4$3p, 2s2p$^3$3s$^2$, 2s2p$^3$3p$^2$, 2s2p$^3$3d$^2$, 2s2p$^3$3s3d, 2p$^5$3s, 2p$^5$3d, 2p$^4$3s3p, 2p$^4$3p3d, 2p$^3$3d$^3$, 2p$^3$3s$^2$3d, 2p$^3$3s3p$^2$, 2p$^3$3p$^2$3d, 2p$^3$3s3d$^2$ odd-parity configurations were explicitly included in the HFR physical model. 
Core-valence correlations were modeled by a core-polarization potential corresponding to a He-like O VII ionic core with a dipole polarizability $\alpha$$_d$ = 0.0026 a$_0^3$ \citep{johnson} and a cut-off radius $r_c$ = 0.198 a$_0$. The latter value corresponds to the calculated 
HFR value of $<$1s$|r|$1s$>$ of the outermost core orbital (1s). In addition, a semi-empirical fitting of the energy levels was performed in order to reduce the discrepancies between the calculated and experimental wavelengths for the three transitions of interest.

Figure~\ref{OAdipole} depicts our HFR+CPOL $gf$-value by a long-dashed horizontal line. This lays right in between the 
length and velocity gauges obtained with {\sc AUTOSTRUCTURE} and the 
Babushkin (B) and Coulomb (C) gauges obtained with MCDHF.

\subsubsection{Nitrogen}

For the calculations in neutral nitrogen, the following configurations were included in the HFR model : 2s$^2$2p$^3$, 2s$^2$2p$^2$$n$p ($n$ = 3--6), 2s$^2$2p$^2$$n$f ($n$ = 4--6), 2s$^2$2p$^2$6h, 2s$^2$2p3s$^2$, 2s$^2$2p3p$^2$, 2s$^2$2p3d$^2$, 2s$^2$2p3s3d, 2s$^2$3p$^3$, 2s$^2$3s$^2$3p,2s$^2$3p3d$^2$, 2s$^2$3s3p3d, 2s2p$^3$3s, 2s2p$^3$3d, 2s2p$^2$3s3p, 2s2p$^2$3p3d, 2p$^5$, 2p$^4$3p, 2p$^3$3s$^2$, 2p$^3$3p$^2$, 2p$^3$3d$^2$, 2p$^3$3s3d, 2p$^2$3p$^3$, 2p$^2$3s$^2$3p, 2p$^2$3p3d$^2$, 2p$^2$3s3p3d (odd parity), and 2s$^2$2p$^2$$n$s ($n$ = 3--6), 2s$^2$2p$^2$$n$d ($n$ = 3--6), 2s$^2$2p$^2$$n$g ($n$ = 5--6), 2s$^2$2p3s3p, 2s$^2$2p3p3d, 2s$^2$3d$^3$, 2s$^2$3s$^2$3d, 2s$^2$3s3p$^2$, 2s$^2$3p$^2$3d, 2s$^2$3s3d$^2$, 2s2p$^4$, 2s2p$^3$3p, 2s2p$^2$3s$^2$, 2s2p$^2$3p$^2$, 2s2p$^2$3d$^2$, 2s2p$^2$3s3d, 2p$^4$3s, 2p$^4$3d, 2p$^3$3s3p, 2p$^3$3p3d, 2p$^2$3d$^3$, 2p$^2$3s$^2$3d, 2p$^2$3s3p$^2$, 2p$^2$3p$^2$3d, 2p$^2$3s3d$^2$ (even parity). The core-polarization parameters corresponding to a He-like N VI ionic core were chosen as $\alpha$$_d$ = 0.0046 a$_0^3$ \citep{johnson} and $r_c$ = 0.228 a$_0$. As for oxygen, a semi-empirical adjustment of the computed energies was performed in order to reproduce the experimental wavelengths for the two N I lines of interest as well as possible.

The HFR+CPOL results for both lines of interest are shown in Figs.~\ref{Ndipole1}, \ref{Ndipole2} by long-dashed horizontal lines. 
Relative to the $gf$-values found with {\sc AUTOSTRUCTURE} and MCDHF, the HFR+CPOL results are about 10\% below for the $\lambda$~8683 line and $\sim 10\%$ above for the $\lambda$~8629 line.
We consider this level of agreement as satisfactory, given the complexity of these nitrogen transitions. 
However, it is clear the level of uncertanity of the nitrogen lines will be larger than for the oxygen lines.

\section{Multiconfiguration Dirac-Hartree-Fock Calculations}

We also calculated the radiative parameters using the purely relativistic multiconfiguration Dirac-Hartree-Fock (MCDHF) method developed by \citet{grant} and \citet{froese}.  
This is implemented in
{\sc GRASP} (General Relativistic Atomic Structure Program) and we use the latest version of this code, {\sc GRASP2018}
\citep{froese}

Different physical models were applied to each atom in order to optimize the wave functions and the corresponding energy levels by gradually increasing the basis of configuration state functions (CSFs), and thus taking into account more correlations. To begin with,
a multireference (MR) of configurations was chosen as being composed of 2s$^2$2p$^k$, 2s$^2$2p$^{k-1}$3s, 2s$^2$2p$^{k-1}$3p and 2s$^2$2p$^{k-1}$3d with $k$ = 3, 4 for N I and O I, respectively. Then, valence-valence (VV) models, in which SD excitations of valence electrons, i.e. occupied subshells of configurations from the MR except the 1s orbital, to a set of active orbitals were considered in order to generate the CSF expansions. These sets of active orbitals were gradually extended up to 9s, 9p, 9d, 6f, 6g and 6h. In each VV model, we added core-valence (CV) correlations by considering single excitations from the 1s orbital to those characterizing the MR configurations. 
It was found that the CV correlations are negligible (i.e. the results of
the VV and CV models are essentially identical) for the transitions of interest in 
both oxygen and nitrogen. Thus, while we describe both sets of models we only
present the CV model results and refer to them as MCDHF values. 
essentially identical Finally, the high-order relativistic effects, i.e. the Breit interaction,
QED self-energy, and vacuum polarization effects were incorporated in the relativistic configuration interaction 
(RCI) step of the
{\sc GRASP2018} package. The details of the MCDHF calculations carried out in oxygen and nitrogen atoms are given below.

Tables~\ref{oihfrenergy} and \ref{nihfrenergy}
compare the calculated energies of the levels of interest with experimental energies
from NIST \citep{nist}. For calculated energies we take those from the largest models described below. 

\subsection{Oxygen}

In neutral oxygen, the MR consisted in 2s$^2$2p$^4$, 2s$^2$2p$^3$3s, 2s$^2$2p$^3$3p and 2s$^2$2p$^3$3d configurations,
 from which all the levels were considered to optimize the 1s, 2s, 2p, 3s, 3p, and 3d orbitals. A first 
VV model (VV1) was built by adding to the MR configurations, SD excitations from 2s, 2p, 3s, 3p and 3d to the 
\{$nl$,$n'l'$,...\} active set of orbitals, where $n$, $n'$,... stand for the maximum principal quantum numbers corresponding to the $l$, $l'$, ... orbital quantum numbers. In the first VV model (VV1), we considered  \{4s,4p,4d,4f\} as active set. In this step, only the new orbitals were optimized, the other ones being kept to their values obtained before. The same strategy was used to build more elaborate VV models by successively considering the following sets of active orbitals : \{5s,5p,5d,5f,5g\} for VV2, \{6s,6p,6d,6f,6g,6h\} for VV3, \{7s,7p,7d,6f,6g,6h\} for VV4, \{8s,8p,8d,6f,6g,6h\} for VV5, and \{9s,9p,9d,6f,6g,6h\} for VV6. For each of these VV models, core-valence correlations were added by means of single excitations from the 1s orbital to the active orbitals of the MR, giving rise to the CV1, CV2, CV3, CV4, CV5, and CV6 models, gradually increasing the number of CSFs from 40 (obtained for the MR) to 21065, 65240, 145259, 193535, 249311, and 313127, respectively. The convergence of the oscillator strengths for the three O I lines of interest was verified  
by comparing the
results obtained using physical models including increasingly large active sets of orbitals, i.e. from MR model to CV6 calculations, and by observing a good agreement between the $gf$-values computed within the Babushkin (B) and 
Coulomb (C) gauges. Such a convergence of results is shown in Fig.~\ref{OAdipole}. 

\subsection{Nitrogen}

The strategy developed for the MCDHF calculations in atomic nitrogen was exactly the same as for oxygen. In this case, the MR was composed of 2s$^2$2p$^3$, 2s$^2$2p$^2$3s, 2s$^2$2p$^2$3p and 2s$^2$2p$^2$3d configurations, from which VVn and CVn ($n$ = 1--6) models were built using the same active sets as those considered in O I. The number of CSFs was found to be equal to 28 (MR), 8316 (CV1), 24715 (CV2), 53504 (CV3), 75728 (CV4), 92718 (CV5), and 117304 (CV6). Here also, a very good convergence of the oscillator strengths was found when going from MR model to CV6 model and when comparing the Babushkin and Coulomb results for the two N I lines of interest. This convergence is shown in Figs.~\ref{Ndipole1}, \ref{Ndipole2}.
\section{Comparisons with previous results and recommendations}

Table~\ref{oitable1} compares the $gf$-values and transition probabilities for the oxygen lines. All three transitions have the same A-value and we list this value only once. From the present calculations we list only the results of our largest models, as these are expected to be the best converged.

We find that all O calculations converge in good agreement with each other (see Fig.~\ref{OAdipole}).
Our best {\sc AUTOSTRUCTURE} calculations yield very close agreement between the length and velocity gauges of the oscillator strengths, with a difference of only $\sim 6\%$. 
Similar agreement is found between the Babushkin and Coulomb gauges in our own MCDHF  
calculations (i.e. $\sim 9\%$). The R-matrix method yields the most uncertain $gf$-values, 
with a difference between the length and velocity gauges of 15\%. The HFR+CPOL results agree closely with our other calculations. 

Practical applications of the atomic data would require us to assert a single recommended value for each transition. Though, there is no clear justification 
for picking the result of any method employed here over the others. Regarding the different gauges, transitions among low lying levels are generally dominated by the outer part of the wavefunctions and the length or Babushkin gauges are generally preferred over the velocity ones. However, we do not believe the information provided by the velocity gauges should be discarded. Thus, we propose the recommended values should be obtained from some average over the different results. We calculated such averages in various ways, including giving half weight to the velocity gauge or excluding them completely, with and without the most discrepant R-matrix results, and including all results with equal weight. It is found the computed averages are all in close agreement (within a few percent). Hence, we decide to adopt the simplest approach. That is, we 
recommend $gf$-values and transition probabilities obtained by averaging over all results and estimate uncertainties on the values from the observed dispersion. Our results and recommended values for the O I lines of interest are presented in 
Table~\ref{oitable1}. The $\log(gf)$ values are 0.350, 0.196, $-0.030$ dex for the 7771, 7774, and 7775 \AA\  lines, correspondingly, which are very similar to the values adopted by 
\citet{bergemann} and \citet{magg} (0.350, 0.204, $-0.019$ dex), based on {\sc AUTOSTRUCTURE} calculations alone. 
The very minor differences - at a sub-percent level - are caused by the taking into account 
the results from other methods in the averaging procedure in this work. We note that our new $gf$-values in this work are slightly lower than in \citet{bergemann} and \citet{magg}, hence the O solar abundance will be about 0.006 dex higher.
We also tabulate the length and velocity gauges of the \citet{hibbert} calculations using the CIV3 code and the recent values by \citet{civis} using the QDT method. The CIV3 length gauge value is 
recommended by NIST \citep{nist} in their list of transition probabilities. NIST gives a quality rating ``A" to these rates, which is equivalent to an uncertainty of $\sim 3\%$.  We find this rate to be at the upper end, slightly above 
our estimated $1\sigma$ ($\sim 5\%$) error bars. Regarding the values of \citet{civis}, these are 1.5$\sigma$ below our recommended rates.

The $gf$-values and transition probabilities for lines of interest in the N I are presented in Table~\ref{n1comps}. 
The convergence of these transitions is much slower than for the O I lines and even our largest models yield 
significant dispersion among the results. Taking all results into consideration, we arrive to recommended rates for both N I transitions with uncertainties just over 10\%. The NIST dababase adopts transitions rates from \citet{tachiev} and assigns an accuracy rating of "B+" ($<7\%$) for both transitions. We find their $gf$-values to be nearly 10\% below and 5\% above our recommended values for the $\lambda 8683$  and $\lambda 8629$ lines, respectively. The previous calculation by \citet{hibbertb} yielded $f$-values about 3\% higher than in \citet{tachiev}. Experimental measurements of transition probabilities for the lines were reported by \citet{bridges} and \citet{musielok}. We find our recommended rate for the $\lambda 8683$ line is $\sim 1.5\sigma$ larger than the experimental determination. On the other hand, our recommended rate for the $\lambda 8629$ line is in excelent agreement with the \cite{musielok} experimental value.
\section{The oxygen photoionization cross sections and their effect on modeling solar abundances}

\citet{bergemann}, in the analysis of the solar photospheric spectrum, employed newly calculated photoionization cross sections. These calculations were motivated by some shortcomings in the widely-used cross sections from the OPACITY Project (OP) \citep{OP, topbase}. The issues with the OP 
cross sections are: (i) they are only available in LS coupling, (ii) the energy resolution of the cross sections is too 
low to allow for accurate integration over resonances, (iii) only total cross sections, level-to-level partial
cross sections, are available, and (iv) the OP cross sections are only available for states up to $n\le 10$.
It should be noted that our intention is not to criticize the OP data, which serves very well the purpose for which they were computed. The purpose was to provide averaged atomic opacities for the solar interior, where neutral oxygen contributes very little. However, in photospheric non-LTE spectral analysis, the accuracy requirements are much higher and it is important to use the most reliable atomic data.

Unfortunately, we made a mistake in the top panel of Figure~2 in \citet{bergemann} by plotting a cross section different than intended. Specifically, the figure plots our cross section for the second $^5$S$^o$ bound state, which is the 4s~$^5$S$^o$ state, rather than the intended first $^5$S$^o$ bound state, or 3s~$^5$S$^o$. 
This led \citet{nahar} to criticize all atomic data used in \citet{bergemann} and cast doubts on the reliability of the oxygen abundance determination. 

Our photoionization cross sections were computed in intermediate coupling using the
Breit-Pauli version of the code RMATRX \citep{rmatrx}.
We used {\sc AUTOSTRUCTURE} to compute the atomic orbitals for the O$^+$ target ion.
Our target representation included eleven configurations
2s$^2$2p$^3$, 2s$^2$2p$^2$3s, 2s$^2$2p$^2$3p, 2s$^2$2p$^2$3d, 2s2p$^4$, 2s2p$^3$3s, 2s2p$^3$3p,
2s2p$^3$3d, 2s$^2$2p3s$^2$, 2s$^2$2p3p$^2$, and 2s$^2$2p3d$^2$. This expansion led to 26
close coupling LS terms and 38 energy levels. Our calculation yields 409 bound levels for neutral oxygen with $n\le 20$ and angular momentum $j\le 8$. We calculated the photoionization cross sections for all levels with an energy resolution of $3.3\times 10^{-4}$~Ryd up to the highest ionization threshold. This is followed by much coarser energy resolution at larger energies, where no resonances should be found. 

\begin{figure}
\scalebox{0.5}{\includegraphics[angle=-90]{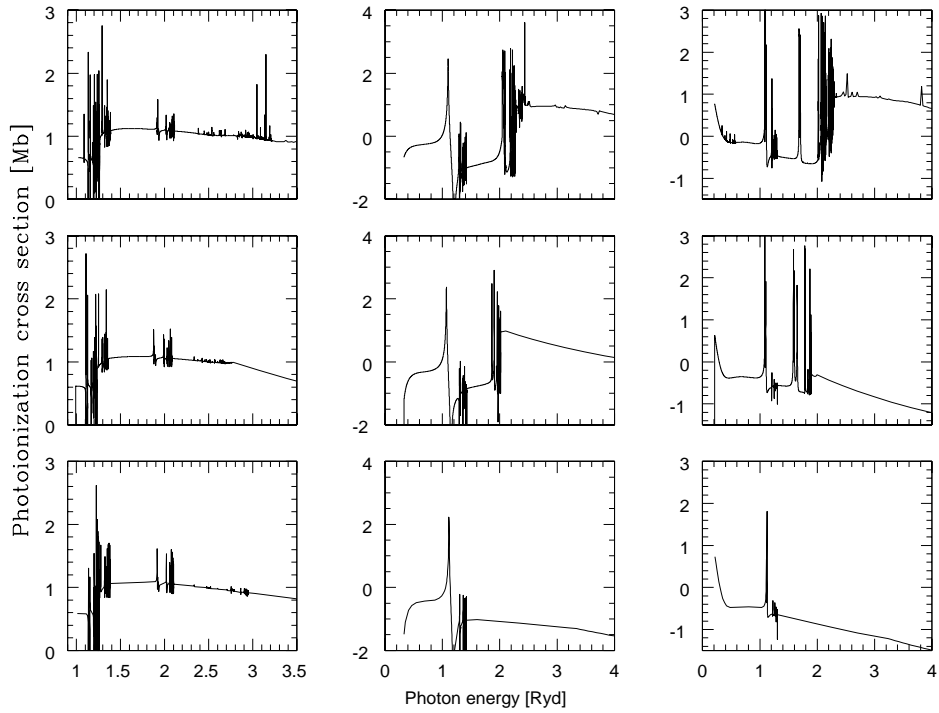}}
  \caption{Photoionization cross sections for the 2s$^2$2p$^4$~$^3$P$_2$ ground level (left panels),
and the 2s$^2$2p$^3$3s~$^5$S$_2^o$ (middle panels), and 2s$^2$2p$^3$3s~$^5$P$_1$ (right panels) levels. 
The results of the present calculation, also reported in \cite{bergemann}, are depicted by the top panels, and 
the results of \cite{nahar21} and the OP \citep{topbase} are presented in the middle and bottom panels, respectively. 
\label{photoxs}}
\end{figure}

Figure~\ref{photoxs} shows the cross sections for the oxygen ground level, 2s$^2$2p$^4$~$^3$P$_2$, 
and the 2s$^2$2p$^3$3s~$^5$S$_2^o$, and 2s$^2$2p$^3$3s~$^5$P$_1$ levels. Here, we also compare with the OP cross section and those from \citet{nahar21}.  By comparing the cross sections for the ground level, we can see all three results are qualitatively very similar. The present cross sections show more resonances in the near threshold region and converge to 2s$^2$2p$^2$3$l$ states of the O$^+$ target. This is  expected, because the present calculations are done in intermediate coupling while previous calculations where in LS-coupling. Our cross section also exhibits resonances for photon energies around 3 Ryd, which are missing in the Nahar and OP calculations. These resonances converge to open 2s configurations of the form 2s2p$^3$3$l$ of O$^+$. Although these configurations are included in Nahar's target description, the states for these configurations must have been omitted from the close coupling expansion.

Contribution from the open 2s configurations are most noticeable in the cross sections for the 2s$^2$2p$^3$3s~$^5$S$^o$ and 2s$^2$2p$^3$3s~$^5$P states for photon energies beyond $\sim 2.4$ Ryd. Coupling to the open sub-shell configurations results in photoionization jumps followed by increases in the cross sections by over two orders of magnitude. These inner-shell effects are clearly missing in the OP cross sections and only partly accounted for in the Nahar cross sections. 

\begin{figure}
\scalebox{0.5}{\includegraphics[angle=-00]{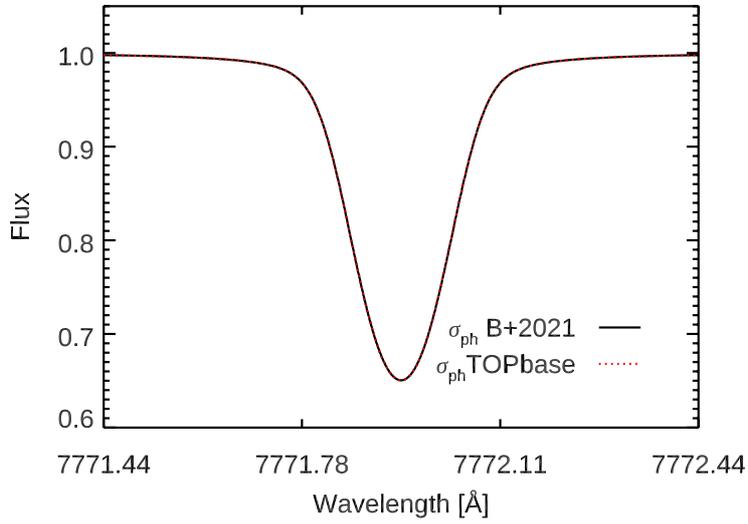}}
  \caption{Model line profile for the O I $\lambda 7771$ transition in the solar photosphere.
The line is modeled using the present photoionization cross sections as in \cite{bergemann} (B+2021)
and the OP cross sections taken from TOPbase \citep{topbase}.
\label{oline}}
\end{figure}

It is worth asking, how critical is the accuracy of the photoionization cross section in the 
analysis of the solar oxygen abundance? The answer to this question is best 
illustrated by Figure~\ref{oline}. The figure compares the non-LTE line profile for 
the O I $\lambda$7771 line as predicted using the present photoionization cross 
sections and the OP cross sections. The model lines are nearly indistinguishable, and the differences in the predicted equivalent widths from the two sets of cross sections for lines of this multiplet are 0.07 m\AA, which corresponds to the equivalent width difference of 0.1\%. The corresponding difference in the O abundance is 0.0009 dex.  This is $30$ times smaller than the error of 0.03~dex in the solar O abundance quoted in \citet{bergemann}. Photoionisation is not important in the non-LTE calculations for O I, because of the very high ionization potential of O I, 13 eV, that renders radiative ionisation processes too inefficient to significantly influence the statistical equilibrium of the ion in the conditions of the solar photosphere.  

\section{Summary and conclusions}

We carried out very large calculations of $f$-values and transition probabilities for dipole 
allowed lines used in NLTE analysis of solar photospheric spectra of oxygen and nitrogen \citep{bergemann, magg}.

We use various methods and codes for these calculations, namely {\sc AUTOSTRUCTURE}, 
R-matrix, {\sc GRASP2018} and {\sc HFR+CPOL}. 

The convergence of calculated oscillator strength values was evaluated by various criteria 
that are the stability of the numeric values as more configurations are accounted for in the CI expansion, the agreement between values calculated by different  gauges, and agreement among different methods.
This allowed us to propose reliable rates with estimated uncertainties. 

We find satisfactory agreement, withing our estimated uncertainties,  between our recommended values with those from NIST \citep{nist}. However, it is seen that the uncertainties 
stated in the NIST database are overestimated. 

We also present the details of R-matrix calculations of oxygen photoionization cross sections 
that were reported in \citet{bergemann}.  Our results are more extensive and detailed than previous calculations by other authors. {We also quantify the influence of adopted photoionization cross sections on the spectroscopic estimate of the solar O abundance, using the OP cross sections. We confirm that the 3D NLTE value presented by Bergemann et al. is robust and 
unaffected by the choice of photoionization data, contrary to the claim made by Nahar (2022)}. 

%We have been able to obtain converged values for the observed lines of nitrogen and oxygen as listed in table~\ref{fvalues}. However, no convergence was achieved for the carbon lines. In the case of carbon the lines available in the spectra are weak and arise from 2s$^2$2p3s and 2s$^2$2p3p, which are difficult to represent accurately. Thus, we see the $f$-values for these lines to vary by several factors for different configuration expansions. In absence of reliable results from our own calculat%ions we adopt the results of \citet{??}, but caution significant uncertainties exist in values derived from these data.

\clearpage

\bibliographystyle{aa}    

\typeout{}  
\bibliography{refs}     

\begin{thebibliography}{37}
\expandafter\ifx\csname natexlab\endcsname\relax\def\natexlab#1{#1}\fi

\bibitem[{{Asplund} {et~al.}(2021){Asplund}, {Amarsi}, \&
  {Grevesse}}]{2021A&A...653A.141A}
{Asplund}, M., {Amarsi}, A.~M., \& {Grevesse}, N. 2021, \aap, 653, A141

\bibitem[{{Badnell}(2011)}]{autoss}
{Badnell}, N.~R. 2011, Computer Physics Communications, 182, 1528

\bibitem[{{Bensby} {et~al.}(2014){Bensby}, {Feltzing}, \& {Oey}}]{bensby}
{Bensby}, T., {Feltzing}, S., \& {Oey}, M.~S. 2014, \aap, 562, A71

\bibitem[{{Bergemann} {et~al.}(2021){Bergemann}, {Hoppe}, {Semenova},
  {Carlsson}, {Yakovleva}, {Voronov}, {Bautista}, {Nemer}, {Belyaev},
  {Leenaarts}, {Mashonkina}, {Reiners}, \& {Ellwarth}}]{bergemann}
{Bergemann}, M., {Hoppe}, R., {Semenova}, E., {et~al.} 2021, \mnras, 508, 2236

\bibitem[{{Bergemann} {et~al.}(2018){Bergemann}, {Sesar}, {Cohen}, {Serenelli},
  {Sheffield}, {Li}, {Casagrande}, {Johnston}, {Laporte}, {Price-Whelan},
  {Sch{\"o}nrich}, \& {Gould}}]{bergemann18}
{Bergemann}, M., {Sesar}, B., {Cohen}, J.~G., {et~al.} 2018, \nat, 555, 334

\bibitem[{{Berrington} {et~al.}(1995){Berrington}, {Eissner}, \&
  {Norrington}}]{rmatrx}
{Berrington}, K.~A., {Eissner}, W.~B., \& {Norrington}, P.~H. 1995, Computer
  Physics Communications, 92, 290

\bibitem[{{Borsa} {et~al.}(2021{\natexlab{a}}){Borsa}, {Fossati}, {Koskinen},
  {Young}, \& {Shulyak}}]{2021NatAs.tmp..253B}
{Borsa}, F., {Fossati}, L., {Koskinen}, T., {Young}, M.~E., \& {Shulyak}, D.
  2021{\natexlab{a}}, Nature Astronomy

\bibitem[{{Borsa} {et~al.}(2021{\natexlab{b}}){Borsa}, {Fossati}, {Koskinen},
  {Young}, \& {Shulyak}}]{2022NatAs...6..226B}
{Borsa}, F., {Fossati}, L., {Koskinen}, T., {Young}, M.~E., \& {Shulyak}, D.
  2021{\natexlab{b}}, Nature Astronomy, 6, 226

\bibitem[{{Bridges} \& {Wiese}(2010)}]{bridges}
{Bridges}, J.~M. \& {Wiese}, W.~L. 2010, \pra, 82, 024502

\bibitem[{{Civi{\v{s}}} {et~al.}(2018){Civi{\v{s}}}, {Kubel{\'\i}k}, {Ferus},
  {Zanozina}, {Pastorek}, {Naskidashvili}, \& {Chernov}}]{civis}
{Civi{\v{s}}}, S., {Kubel{\'\i}k}, P., {Ferus}, M., {et~al.} 2018, \apjs, 239,
  11

\bibitem[{{Collet} {et~al.}(2011){Collet}, {Magic}, \& {Asplund}}]{stagger}
{Collet}, R., {Magic}, Z., \& {Asplund}, M. 2011, in Journal of Physics
  Conference Series, Vol. 328, Journal of Physics Conference Series, 012003

\bibitem[{{Cowan}(1981)}]{cowan}
{Cowan}, R.~D. 1981, {The theory of atomic structure and spectra}

\bibitem[{{Cunto} {et~al.}(1993){Cunto}, {Mendoza}, {Ochsenbein}, \&
  {Zeippen}}]{topbase}
{Cunto}, W., {Mendoza}, C., {Ochsenbein}, F., \& {Zeippen}, C.~J. 1993, \aap,
  275, L5

\bibitem[{{Froese Fischer} {et~al.}(2019){Froese Fischer}, {Gaigalas},
  {J{\"o}nsson}, \& {Biero{\'n}}}]{grasp18}
{Froese Fischer}, C., {Gaigalas}, G., {J{\"o}nsson}, P., \& {Biero{\'n}}, J.
  2019, Computer Physics Communications, 237, 184

\bibitem[{{Froese Fischer} {et~al.}(2016){Froese Fischer}, {Godefroid},
  {Brage}, {J{\"o}nsson}, \& {Gaigalas}}]{froese}
{Froese Fischer}, C., {Godefroid}, M., {Brage}, T., {J{\"o}nsson}, P., \&
  {Gaigalas}, G. 2016, Journal of Physics B Atomic Molecular Physics, 49,
  182004

\bibitem[{{Grant}(2007)}]{grant}
{Grant}, I.~P. 2007, {Relativistic Quantum Theory of Atoms and Molecules},
  Vol.~40

\bibitem[{{Grevesse} \& {Sauval}(1998)}]{grevesse}
{Grevesse}, N. \& {Sauval}, A.~J. 1998, \ssr, 85, 161

\bibitem[{{Gudiksen} {et~al.}(2011){Gudiksen}, {Carlsson}, {Hansteen}, {Hayek},
  {Leenaarts}, \& {Mart{\'\i}nez-Sykora}}]{bifrost}
{Gudiksen}, B.~V., {Carlsson}, M., {Hansteen}, V.~H., {et~al.} 2011, \aap, 531,
  A154

\bibitem[{{Hibbert} {et~al.}(1991{\natexlab{a}}){Hibbert}, {Bi\'emont},
  {Godefroid}, \& {Vaeck}}]{hibbert}
{Hibbert}, A., {Bi\'emont}, E., {Godefroid}, M., \& {Vaeck}, N.
  1991{\natexlab{a}}, Journal of Physics B Atomic Molecular Physics, 24, 3943

\bibitem[{{Hibbert} {et~al.}(1991{\natexlab{b}}){Hibbert}, {Bi\'emont},
  {Godefroid}, \& {Vaeck}}]{hibbertb}
{Hibbert}, A., {Bi\'emont}, E., {Godefroid}, M., \& {Vaeck}, N.
  1991{\natexlab{b}}, \aaps, 88, 505

\bibitem[{{Johnson} {et~al.}(1983){Johnson}, {Kolb}, \& {Huang}}]{johnson}
{Johnson}, W.~R., {Kolb}, D., \& {Huang}, K.~N. 1983, Atomic Data and Nuclear
  Data Tables, 28, 333

\bibitem[{{Kramida} {et~al.}(2021){Kramida}, {Ralchenko}, {Reader}, \& {NIST
  ASD Team}}]{nist}
{Kramida}, A., {Ralchenko}, Y., {Reader}, J., \& {NIST ASD Team}. 2021,
  \[ONLINE\]

\bibitem[{{Lodders}(2019)}]{lodders}
{Lodders}, K. 2019, arXiv e-prints, arXiv:1912.00844

\bibitem[{{Magg} {et~al.}(2022){Magg}, {Bergemann}, {Serenelli}, {Bautista},
  {Plez}, {Heiter}, {Gerber}, {Ludwig}, {Basu}, {Ferguson}, {Gallego},
  {Gamrath}, {Palmeri}, \& {Quinet}}]{magg}
{Magg}, E., {Bergemann}, M., {Serenelli}, A., {et~al.} 2022, \aap, 661, A140

\bibitem[{{Musielok} {et~al.}(1995){Musielok}, {Wiese}, \& {Veres}}]{musielok}
{Musielok}, J., {Wiese}, W.~L., \& {Veres}, G. 1995, \pra, 51, 3588

\bibitem[{{Nahar}(2021)}]{nahar21}
{Nahar}, S.~N. 2021, Atoms, 9, 73

\bibitem[{{Nahar}(2022)}]{nahar}
{Nahar}, S.~N. 2022, \mnras, 512, L39

\bibitem[{{Quinet} {et~al.}(2002){Quinet}, {Palmeri}, {Bi\'emont}, {Li}, \&
  {Svanberg}}]{quinet02}
{Quinet}, P., {Palmeri}, P., {Bi\'emont}, E., {Li}, Z.~S., \& {Svanberg}, S.
  2002, Journal of Alloys and Compounds, 344, 255

\bibitem[{{Quinet} {et~al.}(1999){Quinet}, {Palmeri}, {Bi{\'e}mont}, {McCurdy},
  {Rieger}, {Pinnington}, {Wickliffe}, \& {Lawler}}]{quinet99}
{Quinet}, P., {Palmeri}, P., {Bi{\'e}mont}, E., {et~al.} 1999, \mnras, 307, 934

\bibitem[{{Schuler} {et~al.}(2021){Schuler}, {Andrews}, {Clanzy}, {Mourabit},
  {Chanam{\'e}}, \& {Ag{\"u}eros}}]{schuler}
{Schuler}, S.~C., {Andrews}, J.~J., {Clanzy}, V.~R., {et~al.} 2021, \aj, 162,
  109

\bibitem[{{Seaton}(1987)}]{OP}
{Seaton}, M.~J. 1987, Journal of Physics B Atomic Molecular Physics, 20, 6363

\bibitem[{{Serenelli} {et~al.}(2009){Serenelli}, {Basu}, {Ferguson}, \&
  {Asplund}}]{serenelli2009}
{Serenelli}, A.~M., {Basu}, S., {Ferguson}, J.~W., \& {Asplund}, M. 2009,
  \apjl, 705, L123

\bibitem[{{Serenelli} {et~al.}(2011){Serenelli}, {Haxton}, \&
  {Pe{\~n}a-Garay}}]{serenelli2011}
{Serenelli}, A.~M., {Haxton}, W.~C., \& {Pe{\~n}a-Garay}, C. 2011, \apj, 743,
  24

\bibitem[{{Tachiev} \& {Froese Fischer}(2002)}]{tachiev}
{Tachiev}, G.~I. \& {Froese Fischer}, C. 2002, \aap, 385, 716

\bibitem[{{Tautvai{\v{s}}ien{\.{e}}} {et~al.}(2022){Tautvai{\v{s}}ien{\.{e}}},
  {Drazdauskas}, {Bragaglia}, {Martell}, {Pancino}, {Lardo}, {Mikolaitis},
  {Minkevi{\v{c}}i{\={u}}t{\.{e}}}, {Stonkut{\.{e}}}, {Ambrosch}, {Bagdonas},
  {Chorniy}, {Sanna}, {Franciosini}, {Smiljanic}, {Randich}, {Gilmore},
  {Bensby}, {Bergemann}, {Gonneau}, {Guiglion}, {Carraro}, {Heiter}, {Korn},
  {Magrini}, {Morbidelli}, \& {Zaggia}}]{2022A&A...658A..80T}
{Tautvai{\v{s}}ien{\.{e}}}, G., {Drazdauskas}, A., {Bragaglia}, A., {et~al.}
  2022, \aap, 658, A80

\bibitem[{{Turrini} {et~al.}(2021){Turrini}, {Schisano}, {Fonte}, {Molinari},
  {Politi}, {Fedele}, {Pani{\'c}}, {Kama}, {Changeat}, \&
  {Tinetti}}]{2021ApJ...909...40T}
{Turrini}, D., {Schisano}, E., {Fonte}, S., {et~al.} 2021, \apj, 909, 40

\bibitem[{{Villante} \& {Serenelli}(2020)}]{villante}
{Villante}, F.~L. \& {Serenelli}, A. 2020, arXiv e-prints, arXiv:2004.06365

\end{thebibliography}

\clearpage

\begin{table*}
\caption{CI expansions used in the calculation of radiative rates of neutral oxygen.} 
\label{oiconfigs}
\centering       
\begin{tabular}{c r l}
\hline\hline
{Expan.} & size & {configurations} \\
\hline 
CI-A & 15 & 2s$^2$2p$^4$, 2s$^2$2p$^3nl~(n = 3 \to 4, l = 0 \to 2)$,
, 2s2p$^5$, 2p$^6$, 2s2p$^4 nl~(n = 3 \to 4, l = 0 \to 2)$,  \cr
\hline
CI-B & 30 & CI-A, 2s2p$^3$3s$nl ~(n=3 \to 4, l=0 \to 2)$, 2s2p$^3$3p$^2$, 2s2p$^3$3p3d,
2s2p$^3$3p4$l ~( l=0 \to 2)$, 2s2p$^3$3d$^2$,\cr
 & & 2s2p$^3$3d4$l~(l=0 \to 2)$\cr
\hline
CI-C & 36 & CI-B, 2s2p$^3$4s$4l ~(l=0 \to 2)$, 2s2p$^3$4p$^2$, 2s2p$^3$4p4d, 2s2p$^3$4d$^2$\cr
\hline
CI-D & 21 & CI-A, 2p$^5$$nl ~(n=3 \to 4, l=0 \to 2)$\cr
\hline
CI-E & 30 & CI-A, 2p$^4$3s$nl ~(n=3 \to 4, l=0 \to 2)$, 2p$^4$3p$^2$, 2p$^4$3p3d, 2p$^4$3p4$l~(l=0 \to 2)$, 2p$^4$3d$^2$,\cr
         &     & 2p$^4$3d4$l~(l=0 \to 2)$\cr
\hline
CI-F & 30 & CI-A, 2p$^3$3s$^2$3p, 2p$^3$3s$^2$3d, 2p$^3$3s$^2$4$l~(l=0 \to 2)$,
2p$^3$3s3p$^2$,  2p$^3$3p$^2$3d, 2p$^3$3p$^2$$4l~(l=0 \to 2)$, \cr
        &     &
         2p$^3$3s3d$^2$,  2p$^3$3p3d$^2$, 2p$^3$3d$^2$$4l ~( l=0~-~2)$\cr
\hline
CI-G & 44& CI-B, 2s$^2$2p$^2$3s$nl~(n=3 \to 4, l=0 \to 2)$, 2s$^2$2p$^2$3p$^2$, 2s$^2$2p$^2$3p3d, 2s$^2$2p$^2$3p4$l ~(l=0 \to 2)$,\cr
         &     & 2s$^2$2p$^2$3d$^2$, 2s$^2$2p$^2$3d4$l~(l=0 \to 2)$
\cr
\hline
CI-H & 58& CI-G, 2s2p$^2$3s$^2$3p, 2s2p$^2$3s$^2$3d, 2s2p$^2$3s$^24l~(l=0 \to 2)$,
2s2p$^2$3p$^3$, 2s2p$^2$3p$^2$3d, 2s2p$^2$3d$^3$,\cr
         &    &  2s2p$^2$3p$^2$$4l~(l=0 \to 2)$, 2s2p$^2$3d$^2$$4l~(l=0 \to 2)$\cr
\hline
CI-I & 50 & CI-A, 2s$^2$2p$^3$4f, 2s$^2$2p$^3$$5l ~(l=0 \to 2)$, 2s2p$^4$4f,
2s2p$^4$$5l~(l=0 \to 2)$, 2s$^2$2p$^2$3s4f, 2s$^2$2p$^2$3p$^2$,  \cr
         &     & 2s$^2$2p$^2$3s$nl ~(n = 3 \to 5, l=0 \to 2)$,
2s$^2$2p$^2$3p3d, 2s$^2$2p$^2$3p4f, 2s$^2$2p$^2$3p$nl~(n = 3 \to 5, l=0 \to 2)$, \cr
         &    &
  2s$^2$2p$^2$3d$^2$, 2s$^2$2p$^2$3d$nl ~(n=4 \to 5, l=0 \to 2)$,
          2s$^2$2p$^2$3d4f \cr
\hline
CI-J & 106 & CI-I, 2p$^6$, 2p$^3$3s$^2$3p, 2p$^3$3s$^2$3d,
        2p$^3$3s$^2$$nl~(n=4 \to 5, l=0 \to 2)$, 2p$^3$3s$^2$4f, 2p$^3$3s3p$^2$, 2p$^3$3p$^2$3d,\cr
       &         & 2p$^3$3p$^2$$nl~(n=4 \to 5, l=0 \to 2)$, 2p$^3$3p$^2$4f,
2p$^3$3s3d$^2$, 2p$^3$3p3d$^2$, 2p$^3$3d$^2$$nl~(n=4 \to 5, l=0 \to 2)$, \cr
       &         & 2p$^3$3d$^2$4f
        2s$^2$2p3s$^2$3p, 2s$^2$2p3s$^2$3d, 2s$^2$2p3s$^2$$nl~(n=4 \to 5, l=0 \to 2)$,
2s$^2$2p3s$^2$4f, 2s$^2$2p3p$^3$, \cr
       &         & 2s$^2$2p3p$^2$3d, 2s$^2$2p3p$^2$$nl~(n=4 \to 5, l=0 \to 2)$, 2s$^2$2p3p$^2$4f,
2s$^2$2p3d$^3$, 2s$^2$2p3s3d$^2$, 2s$^2$2p3p3d$^2$, \cr
       &         & 2s$^2$2p3d$^2$$nl~(n=4 \to 5, l=0 \to 2)$, 2s$^2$2p3d$^2$4f\cr
\hline  
\end{tabular}
\end{table*}

\begin{table*}
\caption{Comparison of experimental and predicted {\it ab initio} energies in Ryd from our
different AUTOSTRUCTURE target expansions for oxygen.} 
\label{oiener}
\centering       
\begin{tabular}{lllllllllllll}
\hline\hline
{Conf.} &{Term}&{Exp.} & {CI-A} &{CI-B} &{CI-C} &{CI-D}
&{CI-E} &{CI-F} &{CI-G}
&{CI-H} &{CI-I} &{CI-J} \\
\hline
2p$^4$  &$^3$P  &0.0000 &0.0000 &0.0000 &0.0000 &0.0000 &0.0000 &0.0000 &0.0000 &0.0000 &0.0000 &0.0000 \cr
2p$^4$  &$^1$D  &0.1425 &0.1603 &0.1607 &0.1601 &0.1595 &0.1601 &0.1603 &0.1610 &0.1611 &0.1592 &0.1601 \cr
2p$^4$  &$^1$S  &0.3059 &0.3470 &0.4438 &0.4455 &0.3351 &0.4424 &0.4432 &0.4436 &0.4436 &0.4403 &0.3360 \cr
2p$^3$3s&$^5$S$^o$  &0.6702 &0.6564 &0.5895 &0.6026 &0.6675 &0.6615 &0.6568 &0.5990 &0.5968 &0.6561 &0.6555 \cr
2p$^3$3s&$^3$S$^o$  &0.6977 &0.6904 &0.6256 &0.6375 &0.7005 &0.6949 &0.6911 &0.6347 &0.6352 &0.6870 &0.6878 \cr
2p$^3$3p&$^5$P&0.7874&0.7726&0.7036 &0.7152 &0.7823 &0.7765 &0.7728 &0.7131 &0.7131 &0.7870 &0.7751 \cr
2p$^3$3p&$^3$P&0.8056&0.7931&0.7259 &0.7381 &0.8036 &0.7988 &0.7935 &0.7380 &0.7379 &0.8229 &0.7989 \cr
2p$^3$4s&$^5$S$^o$  &0.8680 &0.8506 &0.8554 &0.7946 &0.8607 &0.8558 &0.8510 &0.8093 &0.8095 &0.8619 &0.8554 \cr
2p$^3$4s&$^3$S$^o$  &0.8748 &0.8598 &0.8655 &0.8050 &0.8696 &0.8660 &0.8603 &0.8099 &0.8101 &0.8691 &0.8639 \cr
2p$^3$3d&$^5$D$^o$  &0.8857 &0.8677 &0.7992 &0.8272 &0.8776 &0.8719 &0.8681 &0.8711 &0.8710 &0.8699 &0.8698 \cr
2p$^3$3d&$^3$D$^o$  &0.8863 &0.8685 &0.7999 &0.8305 &0.8783 &0.8726 &0.8689 &0.8791 &0.8778 &0.8699 &0.8706 \cr
2p$^3$3s&$^3$D$^o$  &0.9197 &0.9407 &0.8751 &0.8912 &0.9549 &0.9486 &0.9414 &0.8808 &0.8807 &0.9375 &0.9352 \cr
2p$^3$3p&$^3$P&0.8056&1.0634&0.9974 &1.0130 &1.0774 &1.0718 &1.0640 &1.0064 &1.0064 &1.0931 &1.0645 \cr
2p$^3$3d&$^3$D  &0.8863 &1.1425 &1.0753 &1.1075 &1.1558 &1.1497 &1.1432 &1.0826 &1.0830 &1.1441 &1.1435 \cr
2p$^3$3p&$^3$P&0.8056&1.2409&1.2128 &1.2310 &1.1738 &1.2508 &1.2417 &1.2133 &1.2133 &1.2592 &1.2342 \cr
2p$^3$3d&$^3$D$^o$  &0.8863 &1.3259 &1.2978&1.3339 &1.2577 &1.3354 &1.3268 &1.2988 &1.2991 &1.3243 &1.3210 \cr
\hline  
\end{tabular}
\end{table*}

\begin{table*}
\caption{CI expansions used in the calculation of radiative rates of neutral nitrogen.} 
\label{niconfigs}
\centering
\begin{tabular}{lrl}
\hline\hline
{Expan.} & size & {configurations} \\
\hline     
CI-A & 14 & 2s$^2$2p$^3$, 2s$^2$2p$^2nl\ (n = 3 \to 4, l = 0 \to 2)$, 2s2p$^4$
          2s2p$^3nl\ (n = 3 \to 4, l = 0 \to 2)$\cr
\hline
CI-B & 29 & CI-A, 2s$^2$2p3s$nl\ (n = 3 \to 4, l = 0 \to 2)$, 2s$^2$2p3p$nl\ (n = 3 \to 4, l = 0 \to 2)$\cr
         & & 2s$^2$2p3d$nl\ (n = 3 \to 4, l = 0 \to 2)$\cr
\hline
CI-C & 36 & CI-B, 2p$^5$, 2p$^4nl\ (n = 3 \to 4, l = 0 \to 2)$\cr
\hline
CI-D & 51 & CI-C, 2s2p$^2$3s$nl ~(n=3\to 4, l=0 \to 2)$, 2s2p$^2$3p$nl ~(n=3 \to 4, l=0 \to 2)$\cr
         &    &  2s2p$^2$3d$nl ~(n=3 \to 4, l=0 \to 2)$
\cr
\hline
CI-E & 74 & CI-D, 2p$^2$3s$^2$3p, 2p$^2$3s$^2$3d, 2p$^2$3s$^24l ~(l=0 \to 2)$, 2p$^2$3s3p$^2$, 2p$^2$3s3p3d, 2p$^2$3s3d$^2$,\cr
         & &  2p$^2$3s3p4$l~(l=0 \to 2)$,
           2p$^2$3s3d4$l~(l=0 \to 2)$, 2p$^2$3p$^3$,
           2p$^2$3p$^2$3d, 2p$^2$3p$^2$4$l~(l=0 \to 2)$,\cr
\hline
CI-F & 126 & CI-E, 2s$^2$2p$^2 5l\ (l = 0 \to 2)$, 2s$^2$2p$^2$4f, 2s2p$^3 5l \ (l = 0 \to 2)$,
2s2p$^3$ 4f , 2s$^2$2p3s5$l\ (l = 0 \to 2)$, 2p$^4$4f, \cr
         & & 2s$^2$2p3s4f, 2s$^2$2p3p5$l\ (l = 0 \to 2)$, 2s$^2$2p3p4f, 2s$^2$2p3d5$l\ (l = 0 \to 2)$, 2s$^2$2p3d4f,\cr
         & & 2p$^4$5$l~(l = 0 \to 2)$, 2s2p$^2$3s5$l \ (l = 0 \to 2)$, 2s2p$^2$3s4f,
2s2p$^2$3p5$l~(l = 0 \to 2)$, 2s2p$^2$3p4f,\cr
         & & 2s2p$^2$3d5$l \ (l = 0 \to 2)$, 2s2p$^2$3d4f,
          2s2p3s$^2 nl\ (n = 3 \to 5, l = 0 \to 2)$, 2s2p3s$^2$4f, 2s2p3s3p$^2$,\cr
         & & 2s2p3s3p3d, 2s2p3s3p$ nl\ (n = 4 \to 5, l = 0 \to 2)$, 2s2p3s3p4f,
2s2p3s3d$^2$, 2s2p3s3d4f,\cr
         & & 2s2p3s3d$ nl~(n = 4 \to 5, l = 0 \to 2)$, 2s2p3p3d$^2$, 2s2p3p3d$ nl~(n = 4 \to 5, l = 0 \to 2)$, 2s2p3p3d4f,\cr
         & & 2s2p3p$^3$, 2s2p3p$^2$3d, 2s2p3p$^2 nl\ (n = 4 \to 5, l = 0 \to 2)$, 2s2p3p$^2$4f\cr
\hline
CI-G & 355& CI-D, 2s$^2$2p$^2$$nl~(n = 5 \to 6, l=0~-~2)$, 2s$^2$2p$^2$4f,  2s$^2$2p$^2$5f, 2s2p$^3$$nl~(n = 5 \to 6, l=0~-~2)$,\cr
       & &  2s2p$^3$4f,  2s2p$^3$5f , 2p$^4$$nl~(n = 5 \to 6, l=0~-~2)$,2s$^2$2p3s$nl~(n = 5 \to 6, l=0~-~2)$, 2p$^4$4f,\cr
         &     & 2p$^4$5f, 2s$^2$2p3s4f,  2s$^2$2p3s5f,2s$^2$2p3p$nl~(n = 5 \to 6, l=0~-~2)$, 2s$^2$2p3p4f,  2s$^2$2p3p5f, 2s$^2$2p3d4f, \cr
&& 2s$^2$2p3d$nl~(n = 5 \to 6, l=0 \to 2)$, 2s$^2$2p3d5f,
2s$^2$2p4s$nl~(n = 4 \to 6, l=0 \to 2)$, 2s$^2$2p4s4f,\cr
&& 2s$^2$2p4s5f,2s$^2$2p4p$^2$, 2s$^2$2p4p4d,
 2s$^2$2p4p$nl~(n = 5 \to 6, l=0~-~2)$, 2s$^2$2p4p4f, 2s$^2$2p4d$^2$,\cr
 & & 2s$^2$2p4d4f, 2s$^2$2p4d5f,
 2s$^2$2p4d$nl~(n = 5 \to 6, l=0~-~2)$, 2s$^2$2p5s4f, 2s$^2$2p5s5f,\cr
 & & 2s$^2$2p5s$nl~(n = 5 \to 6, l=0~-~2)$, 2s$^2$2p5p6$l~(l=0 \to 2)$,\cr
&& 2s$^2$2p5p4f, 2s$^2$2p5p5f, 2s$^2$2p5d5f,
2s$^2$2p5p$^2$, 2s$^2$2p5p5d, 2s$^2$2p5p6$l~(l=0 \to 2)$, 2s$^2$2p5p5f,2s$^2$2p5d$^2$,\cr
&& 2s$^2$2p5d6$l~(l=0 \to 2)$, 2s$^2$2p5d5f,
2s$^2$2p6s6$l~(l=0 \to 2)$, 2s$^2$2p6s5f, 2s$^2$2p6p$^2$, 2s$^2$2p6p6d,\cr
&& 2s$^2$2p6p5f, 2s$^2$2p5f$^2$, 2s2p$^2$3s$nl~(n = 5 \to 6, l=0~-~2)$, 2s2p$^2$3s$nl~(n = 5 \to 6, l=0~-~2)$,\cr
&& 2s2p$^2$3s4f, 2s2p$^2$3s5f, 2s2p$^2$3p4f, 2s2p$^2$3d$nl~(n = 5 \to 6, l=0~-~2)$, 2s2p$^2$3d4f, 2s2p$^2$3d5f,\cr
&& 2s2p$^2$3p5f, 2s2p$^2$4s$nl~(n = 4 \to 6, l=0~-~2)$, 2s2p$^2$4s4f, 2s2p$^2$4s5f,
2s2p$^2$4p$^2$, 2s2p$^2$4p4d, 2s2p$^2$4p4f, \cr
&& 2s2p$^2$4p$nl~(n = 5 \to 6, l=0~-~2)$, 2s2p$^2$4p5f, 2s2p$^2$4d$^2$, 2s2p$^2$4d4f, 2s2p$^2$4d5f\cr
&& 2s2p$^2$4d$nl~(n = 5 \to 6, l=0~-~2)$,
2s2p$^2$5s$nl~(n = 5 \to 6, l=0~-~2)$, 2s2p$^2$5s4f, 2s2p$^2$5s5f, 2s2p$^2$5p4f, \cr
&& 2s2p$^2$5d4f, 2s2p$^2$4f6$l~(l=0~-~2)$, 2s2p$^2$4f5f, 2s2p$^2$5p$^2$, 2s2p$^2$5p5d, 2s2p$^2$5p6$l~(l=0~-~2)$,\cr
&& 2s2p$^2$5p5f, 2s2p$^2$5d$^2$, 2s2p$^2$5d6$l~(l=0~-~2)$, 2s2p$^2$5d5f, 2s2p$^2$6s6$l~(l=0~-~2)$,
2s2p$^2$6s5f,\cr
&& 2s2p$^2$6p$^2$, 2s2p$^2$6d$^2$, 2s2p$^2$6d5f, 2p$^3$3s$nl~(n = 4 \to 6, l=0~-~2)$
2p$^3$3s4f, 2p$^3$3s5f, \cr
& & 2p$^3$3p$^2$, 2p$^3$3p3d, 2p$^3$3p$nl~(n = 4 \to 6, l=0~-~2)$, 2p$^3$3p4f, 2p$^3$3p5f, 2p$^3$3d$^2$, \cr
& & 2p$^3$3d$nl~(n = 4 \to 6, l=0 \to 2)$, 2p$^3$4s4f, 2p$^3$4s5f, 2p$^3$4p$^2$, 2p$^3$4p4d,\cr
& & 2p$^3$4s$nl~(n = 4 \to 6, l=0 \to 2)$, 2p$^3$4p$nl~(5 = 4 \to 6, l=0 \to 2)$, 2p$^3$4p4f, 2p$^3$4p5f, 2p$^3$4d$^2$,\cr
& & 2p$^3$4d$nl~(5 = 4 \to 6, l=0 \to 2)$, 2p$^3$4d4f, 2p$^3$4d5f, 2p$^3$5s$nl~(5 = 4 \to 6, l=0~-~2)$, 2p$^3$5s4f,\cr
& & 2p$^3$5s5f, 2p$^3$5p4f, 2p$^3$5d4f, 2p$^3$4f6$l~(l=0 \to 2)$, 2p$^3$4f5f, 2p$^3$5p$^2$, 2p$^3$5p5d,  \cr
& & 2p$^3$5p6$l~(l=0~-~2)$, 2p$^3$5p5f, 2p$^3$5d$^2$, 2p$^3$5d6$l~(l=0~-~2)$, 2p$^3$6d$^2$, 2p$^3$6d5f\cr
& & 2p$^3$5d5f, 2p$^3$6s6$l~(l=0~-~2)$, 2p$^3$6s5f, 2p$^3$6p$^2$, 2p$^3$6p6d, 2p$^3$6p5f,\cr
\hline  
\end{tabular}
\end{table*}

\begin{table*}
\caption{Comparison of experimental and predicted {\it ab initio} energies in Ryd from our
different AUTOSTRUCTURE target expansions for nitrogen.} 
\label{niener}
\centering        
\begin{tabular}{llllllllll}
\hline\hline
{Conf.} & {Term}& {Exp.} & {CI-A} &{CI-B} &{CI-C} &{CI-D}
&{CI-E} &{CI-F} &{CI-G}
\\
\hline 
2s$^2$2p$^3$	&	$^4$S$^o$	&	0.0000	&	0.0000	&	0.0000	&	0.0000	&	0.0000	&	0.0000	&	0.0000	&	0.0000	\cr

2s$^2$2p$^3$	&	$^2$D$^o$	&	0.1753	&	0.2179	&	0.2143	&	0.2126	&	0.2107	&	0.2111	&	0.2111	&	0.2099	\cr

2s$^2$2p$^3$	&	$^2$P$^o$	&	0.2628	&	0.3488	&	0.3430	&	0.2769	&	0.2860	&	0.2868	&	0.2868	&	0.2864	\cr

2s$^2$2p$^2$3s	&	$^4$P	&	0.7597	&	0.6783	&	0.6902	&	0.6500	&	0.6718	&	0.6720	&	0.6720	&	0.6789	\cr

2s$^2$2p$^2$3s	&	$^2$P	&	0.7857	&	0.7067	&	0.7173	&	0.6771	&	0.6993	&	0.6997	&	0.6997	&	0.7094	\cr

2s2p$^4$	&	$^4$P		&	0.8035	&	0.7991	&	0.7966	&	0.7934	&	0.7754	&	0.7765	&	0.7765	&	0.7757	\cr

2s$^2$2p$^2$3p	&$^2$S$^o$	&	0.8528	&	0.7637	&	0.7775	&	0.7390	&	0.7620	&	0.7614	&	0.7614	&	0.7674	\cr

2s$^2$2p$^2$3p	&$^4$D$^o$	&	0.8646	&	0.7770	&	0.7916	&	0.7515	&	0.7758	&	0.7750	&	0.7750	&	0.7809	\cr

2s$^2$2p$^2$3p	&$^4$P$^o$	&	0.8706	&	0.7845	&	0.7995	&	0.7580	&	0.7828	&	0.7819	&	0.7819	&	0.7877	\cr

2s$^2$2p$^2$3p	&$^4$S$^o$	&	0.8817	&	0.8115	&	0.8265	&	0.7879	&	0.8167	&	0.8149	&	0.8149	&	0.8137	\cr

2s$^2$2p$^2$3p	&$^2$D$^o$	&	0.8827	&	0.8127	&	0.8274	&	0.7879	&	0.8144	&	0.8131	&	0.8131	&	0.8133	\cr

2s$^2$2p$^2$3p	&$^2$P$^o$	&	0.8913	&	0.8269	&	0.8411	&	0.7989	&	0.8263	&	0.8251	&	0.8251	&	0.8230	\cr

2s$^2$2p$^2$3s	&$^2$D	&	0.9082	&	0.8441	&	0.8566	&	0.8164	&	0.8387	&	0.8389	&	0.8389	&	0.8466	\cr

2s$^2$2p$^2$4s	&$^4$P	&	0.9453	&	0.8584	&	0.8703	&	0.8302	&	0.8605	&	0.8618	&	0.8618	&	1.1514	\cr

2s$^2$2p$^2$4s	&$^2$P	&	0.9497	&	0.8689	&	0.8810	&	0.8395	&	0.8721	&	0.8738	&	0.8738	&	1.0892	\cr

2s$^2$2p$^2$3d	&	$^2$P	&	0.9537	&	0.8641	&	0.8753	&	0.8350	&	0.8628	&	0.8625	&	0.8625	&	0.8651	\cr

2s$^2$2p$^2$3d	&	$^4$F	&	0.9547	&	0.8661	&	0.8772	&	0.8366	&	0.8635	&	0.8633	&	0.8633	&	0.8579	\cr

2s$^2$2p$^2$3d	&	$^2$F	&	0.9557	&	0.8680	&	0.8787	&	0.8382	&	0.8658	&	0.8656	&	0.8656	&	1.0153	\cr

2s$^2$2p$^2$3d	&	$^4$P	&	0.9558	&	0.8699	&	0.8803	&	0.8442	&	0.8679	&	0.8678	&	0.8678	&	0.8855	\cr

2s$^2$2p$^2$3d	&	$^4$D	&	0.9570	&	0.8689	&	0.8799	&	0.8391	&	0.8657	&	0.8656	&	0.8656	&	0.8917	\cr

2s$^2$2p$^2$3d	&	$^2$D	&	0.9581	&	0.8711	&	0.8816	&	0.8409	&	0.8686	&	0.8685	&	0.8685	&	0.8953	\cr

2s$^2$2p$^2$4p	&$^2$S$^o$	&	0.9703	&	0.8820	&	0.8943	&	0.8540	&	0.8843	&	0.8855	&	0.8855	&	0.9080	\cr

2s$^2$2p$^2$4p	&$^4$D$^o$	&	0.9739	&	0.8860	&	0.8988	&	0.8581	&	0.8890	&	0.8902	&	0.8902	&	0.9235	\cr

2s$^2$2p$^2$4p	&$^4$P$^o$	&	0.9754	&	0.8886	&	0.9019	&	0.8605	&	0.8918	&	0.8930	&	0.8930	&	0.9224	\cr

2s$^2$2p$^2$4p	&$^2$D$^o$	&	0.9774	&	0.8980	&	0.9127	&	0.8723	&	0.9023	&	0.9030	&	0.9030	&	0.8977	\cr

2s$^2$2p$^2$4p	&$^4$S$^o$	&	0.9791	&	0.9065	&	0.9234	&	0.8851	&	0.9203	&	0.9207	&	0.9207	&	0.8968	\cr

2s$^2$2p$^2$4p	&$^2$P$^o$	&	0.9808	&	0.9125	&	0.9287	&	0.8849
	&	0.9159	&	0.9163	&	0.9163	&	0.9153	\cr
\hline  
\end{tabular}
\end{table*}

\begin{table}
\caption{Comparison of experimental and predicted energies in Ryd from our R-matrix, HFR, and MCDHF calculations for oxygen.}
\label{oihfrenergy}
\centering
\begin{tabular}{lcccc} 
\hline\hline
Level    &Exp.  &R-matrix&HFR&MCDHF \\
\hline
3s $^{5}$S$_{2}$&0.6722& 0.6303& 0.6722& 0.6514\cr
3p $^{5}$P$_{1}$&0.7894& 0.7533& 0.7894& 0.7678\cr
3p $^{5}$P$_{2}$&0.7894& 0.7534& 0.7894& 0.7678\cr
3p $^{5}$P$_{3}$&0.7894& 0.7535& 0.7895& 0.7680\cr
\hline
\end{tabular}
\end{table}

\begin{table}
\caption{Comparison of experimental and predicted energies in Ryd from our HFR and MCDHF calculations for nitrogen.} 
\label{nihfrenergy}
\centering
\begin{tabular}{lccc}  
\hline\hline
Level    &Exp.  &HFR&MCDHF \\
\hline
3s $^{4}$P$_{3/2}$ &	0.7592&	0.7592&0.7587  \cr
3s $^{2}$P$_{3/2}$ &	0.7857&	0.7857&0.7820 \cr
3s $^{2}$D$_{5/2}$ &	0.8642&	0.8642&0.8599 \cr
3p $^{2}$P$_{3/2}$ &	0.8913&	0.8913&0.8872 \cr
\hline
\end{tabular}
\end{table}

\begin{table*}
\caption{Comparison of gf-values and radiative rates with previously
published data for oxygen dipole transitions.} 
\label{oitable1}
\centering 
\begin{tabular}{lcccc} 
\hline\hline 
{Reference} & \multicolumn{3}{c}{gf-value} & {A-value($\times 10^7$ s$^{-1}$)}
\cr
    & {$\lambda$7771} & {$\lambda$7774} & {$\lambda$7775} & \\
\hline     
Present & & & &\cr
AUTOSS(L) & 2.23 & 1.59 & 0.952 & 3.50 \cr
AUTOSS(V) & 2.14 & 1.52 & 0.914 & 3.36 \cr
R-matrix(L) & 2.12 & 1.51 & 0.905 & 3.33 \cr
R-matrix(V) & 2.43 & 1.74 & 1.04 & 3.81 \cr 
HFR+CPOL & 2.24 & 1.60 & 0.959 & 3.52 \cr
MCDHF(B)     & 2.34 & 1.58 & 0.953 & 3.68 \cr
MCDHF(C)     & 2.14 & 1.53 & 0.914 & 3.36 \cr
\hline
\cite{hibbert}& & & &\cr
CIV3(L)  & 2.35 & 1.68 & 1.01 & 3.69 \cr
CIV3(V)  & 2.15 & 1.54 & 0.923 & 3.38 \cr
\cite{civis} & & & & \cr
QDT & 2.07 & 1.48 & 0.889 & 3.25 \cr
\hline
Recommended 
 &$2.24\pm 0.11$&$1.57\pm0.08$&$0.934\pm0.0.046$&$3.53\pm 0.18$\cr
\hline  
\end{tabular}
\end{table*}

%\begin{deluxetable}{ccccccc}
%\tabletypesize{\scriptsize}
%\tablecaption{Comparison of radiative rates with previously
%published data for the oxygen forbidden transition. \label{o1table2}}
%\tablewidth{0pt}
%\tablehead{
%\colhead{AUTOSS} & \colhead{HFR+CPOL} & \colhead{MCDHF} & \colhead{\cite{storey}} &
%\colhead{\cite{hbgv}} &\colhead{\cite{frose}}
%}
%\startdata
%1.0E-2 &   &   &  6.5e-3 & 6.7e-3 & 7.0e-3\cr
%\enddata
%\end{deluxetable}

\begin{table*}
\caption{Comparison of gf-values and radiative rates with previously
published data for nitrogen dipole transitions.} 
\label{n1comps}
\centering
\begin{tabular}{lcccc}
\hline\hline 
{Reference} & \multicolumn{2}{c}{gf-value} & \multicolumn{2}{c}{A-value($\times 10^7$ s$^{-1}$)}\cr
& {$\lambda$8683} & {$\lambda$8629} & {$\lambda$8683} & {$\lambda$8629}
\\
\hline    
Present & & & & \cr
AUTOSS(L) & 1.40 & 1.16 & 2.07 & 2.61 \cr
AUTOSS(V)& 1.23 & 0.904 & 1.82 & 2.03\cr
HFR+CPOL & 1.51 & 1.09  & 2.22 & 2.45 \cr
MCDHF(B) & 1.43 & 1.18  & 1.97 & 2.64 \cr
MCDHF(C) & 1.31 & 1.22  & 1.80 & 2.72 \cr
\hline
\cite{tachiev} & 1.27 & 1.19 & 1.88 & 2.67 \cr
\cite{hibbert} & 1.30 & 1.23 & 1.93 & 2.75 \cr
\cite{bridges} &$1.19\pm 0.10$&     &$1.73\pm 0.14$& \cr
\cite{musielok} &$1.15\pm 0.13$&$1.14\pm0.14$  &$1.67\pm0.19$&$2.55\pm 0.31$\cr 
\hline
Recommended 
 &$1.38\pm 0.15$&$1.14\pm0.12$&$2.03\pm0.0.22$&$2.55\pm 0.27$\cr
\hline   
\end{tabular}
\end{table*}

\end{document}